\documentclass[a4paper,11pt]{article}
\usepackage{pos}

\title{Metastable strings and grand unification}

\author*{Wilfried Buchm\"uller}

\affiliation{Deutsches Elektronen Synchrotron,\\
  Notkestrasse 85, Hamburg, Germany}

\emailAdd{wilfried.buchmueller@desy.de}

\abstract{
The structure of the Standard Model (SM) of particle physics points toward grand unified
theories (GUTs) where strong and electroweak interactions are unified
in a non-Abelian GUT group. The spontaneous breaking of the GUT
symmetry to the SM symmetry, together with cosmic inflation,
generically leads to metastable topological defects, the most
prominent example being cosmic strings. The gravitational-wave
background (GWB) produced by a cosmic string network is one of the
candidates for an explanation of the GWB recently observed by pulsar
timing array (PTA) experiments. We review some properties of the
predicted GWB with emphasis on potential implications for GUT model
building. The most striking prediction is a GWB in the
LIGO-Virgo-KAGRA band that could be discovered in the near future.
}

\FullConference{International Conference  on  Particle Physics and Cosmology (ICPPCRubakov2023)\\
 02-07, October 2023\\
Yerevan, Armenia\\}

\begin{document}
\maketitle

\section{Ultraviolet completion of the Standard Model}

The structure of the Standard Model of particle physics points toward 
``grand unification'' of strong and electroweak interactions: quarks and
leptons form complete multiplets of grand unified (GUT) groups, and the
gauge couplings of strong and electroweak interactions unify at a large energy
scale (GUT scale), approximately without supersymmetry and more precisely
with supersymmetry. The Pati-Salam \cite{Pati:1973uk} and $\text{SO}(10)$ \cite{Georgi:1974my,Fritzsch:1974nn}
GUT groups contain $B\!-\!L$, the difference between baryon and lepton number,
as a local symmetry. The spontaneous breaking of $B\!-\!L$ can generate large
Majorana masses for right-handed neutrinos, and if Yukawa couplings in the
neutrino sector have a similar structure as Yukawa couplings of quarks and
charged leptons, the right order of magnitude of the light-neutrino mass scale
emerges as consequence of the seesaw mechanism. Supersymmetric GUTs
can be derived from higher-dimensional field theories and string theory,
providing a UV completion of the Standard Model
(for recent reviews, see, e.g. \cite{Raby:2017ucc,Hebecker:2021egx}).
From gauge coupling unification we know the order of magnitude of the GUT scale,
$\Lambda_\text{GUT} \simeq 10^{15} \ldots 10^{16}~\text{GeV}$. Since no
evidence for superparticles has been found at the LHC, the scale of
supersymmetry breaking has to lie above a $\text{TeV}$, as for instance in $\text{PeV}$-scale supersymmetry \cite{Wells:2004di,Bhattiprolu:2023lfh}.

\begin{figure}[t]
\begin{center}
\includegraphics[width=0.6\textwidth]{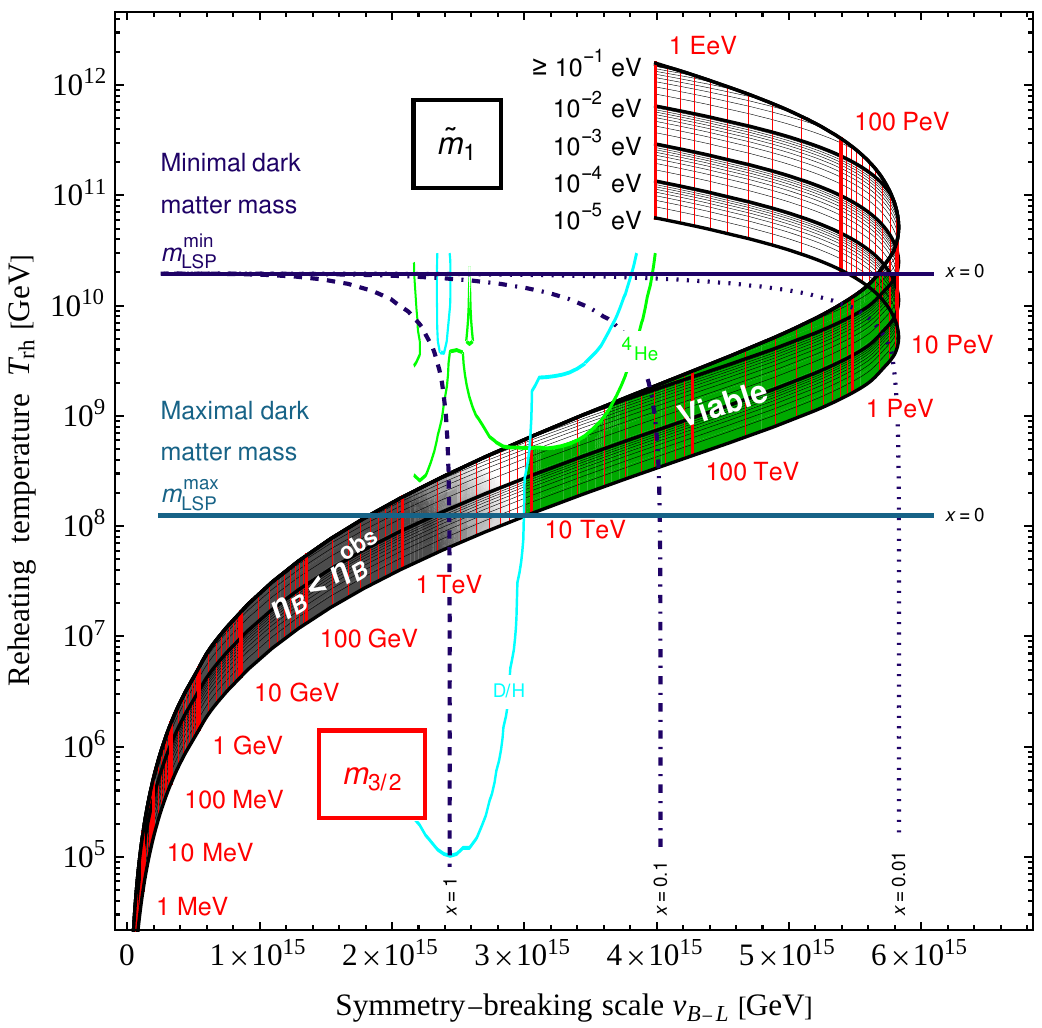}
\caption{Viable parameter space (green) for hybrid inflation, leptogenesis, neutralino DM, and big bang nucleosynthesis.
Hybrid inflation and the dynamics of reheating correlate the parameters $v_{B-L}$, $T_\text{rh}$, $m_{3/2}$ and $\widetilde{m}_1$ (black curves).
Successful leptogenesis occurs outside the gray-shaded region.
Neutralino dark matter is viable in the green region, corresponding to a higgsino (wino) with mass $100 \leq m_{\rm LSP}/\textrm{GeV} \leq 1060$ (2680). From \cite{Buchmuller:2019gfy}.}
\vspace*{0.3cm}
\label{fig:overlay}
\end{center}
\end{figure}

Supersymmetric GUT models contain all ingredients needed in early universe
cosmology: the observed matter-antimatter asymmetry can be generated by
sphaleron processes \cite{Kuzmin:1985mm} from a lepton asymmetry produced in decays of heavy Majorana neutrinos \cite{Fukugita:1986hr}; the lightest superparticle,
gravitino, wino or higgsino, is a candidate for dark matter \cite{Ellis:1983ew};
inflation can be realized in several ways, in particular as hybrid inflation
\cite{Copeland:1994vg,Dvali:1994ms}. A specific example that illustrates the
interplay of these mechanisms is the decay of a false  $B\!-\!L$ vacuum,
broken close to the GUT scale and ending in tachyonic preheating \cite{Felder:2000hj}
where the initial conditions of the hot early universe are generated \cite{Buchmuller:2012wn}. The relevant energy scales are displayed in Fig.~\ref{fig:overlay}: the effective neutrino mass $\tilde{m}_1$ controls leptogenesis, the gravitino mass $m_{3/2}$ provides a constant slope in the almost flat inflaton potential, $v_{B\!-\!L}$ is the scale of $B\!-\!L$ breaking, and $T_{\text{rh}}$ is the reheating
temperature; inflation requires a small Yukawa coupling in the $B\!-\!L$ Higgs
superpotential, leading to a reheating temperatures $T_{\text{rh}} \simeq 10^8 \ldots 10^{10}~\text{GeV}$, and the observed tensor-to-scalar ratio is obtained for gravitino masses in the range $m_{3/3} \simeq 10~\text{TeV} \ldots 10~\text{PeV}$.

During the $B\!-\!L$ phase transition following hybrid inflation a network of cosmic
strings is formed that acts as source of a gravitational wave background
(GWB) (for reviews and references, see, e.g. \cite{Vilenkin:2000jqa,Hindmarsh:2011qj}).
GUT-scale strings have a string tension in the range $G\mu \simeq 10^{-8}\ldots 10^{-6}$, where $G$ denotes Newton's constant and $\mu$ is the energy per unit
length of the string. Stable GUT-scale strings
are excluded by pulsar timing array (PTA)
experiments~\cite{Arzoumanian:2018saf,Kerr:2020qdo,Shannon:2015ect},
whereas strings below the GUT scale are possible and may even render the 
GWB a probe of thermal leptogenesis \cite{Dror:2019syi}.
Moreover, as shown in \cite{Buchmuller:2019gfy}, the PTA bound
on the string tension
can be avoided for metastable cosmic strings that decay into string segments connecting monopole--antimonopole pairs~\cite{Vilenkin:1982hm}. In the semiclassical approximation, the decay rate per string unit length is given by~\cite{Preskill:1992ck,Leblond:2009fq,Monin:2008mp}
\begin{equation}\label{decayrate}
\Gamma_d \simeq \frac{\mu}{2 \pi} \exp\left( - \pi \kappa \right)\ , \quad  
\kappa = \frac{m_M^2}{\mu}  \ ,
\end{equation}
where $m_M$  is the monopole mass. Since the decay rate of a string is proportional
to its length,
\begin{figure}[h!]
\begin{center}
\includegraphics[width=0.6\textwidth]{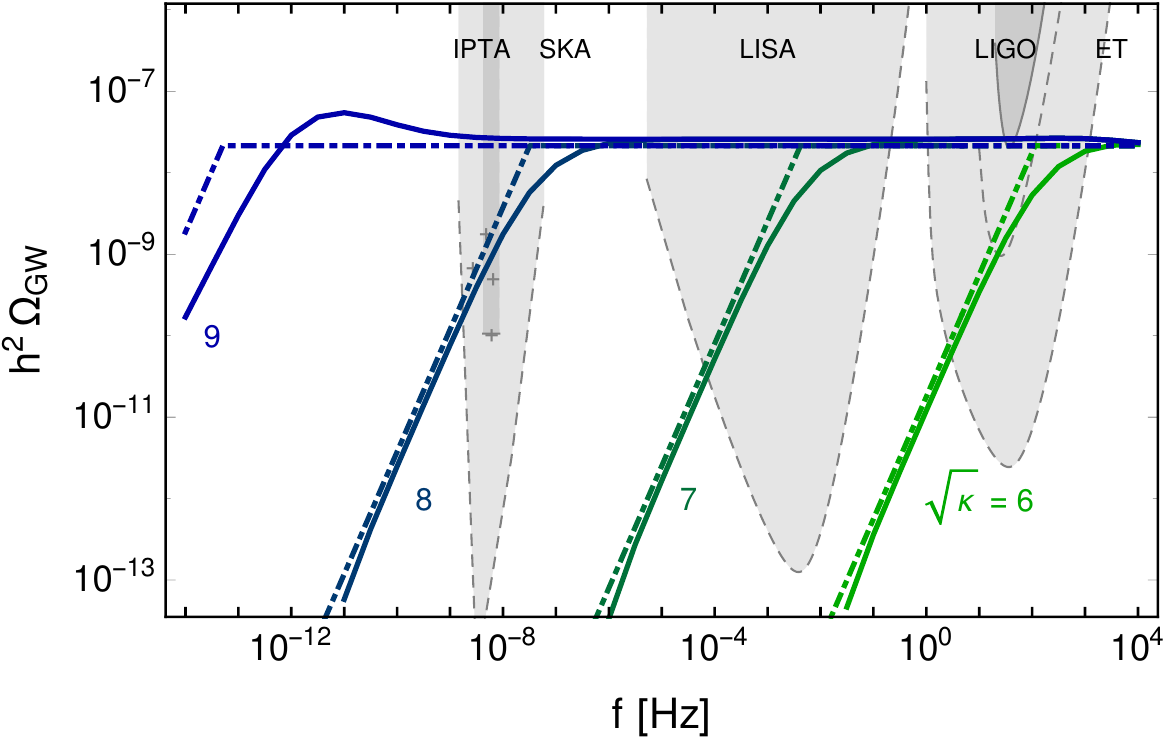}
\caption{GW spectrum for $G \mu = 2 \times 10^{-7}$.
Different values of $\sqrt{\kappa}$ are indicated in different colors;
the blue curve corresponds to a cosmic-string network surviving until
today; the number of SM degrees of freedom is fixed to its
high-temperature value, $g_* =106.75$;
the dot-dashed lines depict an analytical estimate.
The (lighter) gray-shaded areas indicate the sensitivities of (planned) experiments SKA~\cite{Smits:2008cf}, LISA~\cite{Audley:2017drz}, LIGO~\cite{LIGOScientific:2019vic} and ET~\cite{ET}, the crosses within the SKA band indicate constraints by the IPTA~\cite{Verbiest:2016vem}. From \cite{Buchmuller:2019gfy}.}
\label{fig:GWspectrum}
\end{center}
\end{figure}
the spectrum of radiated GWs is suppressed at small frequencies. This
effect is clearly visible in Fig.~\ref{fig:GWspectrum}, and a quantitative analysis shows
that the PTA bound can be avoided for values $\kappa \lesssim 8$.

\section{Gravitational-wave background}
\label{sec:sgwb}
Three years ago the NANOGrav collaboration reported evidence for a stochastic common-spectrum process at nanohertz frequencies~\cite{Arzoumanian:2020vkk}.
The results received considerable attention since the signal could be interpreted
as a GWB of astrophysical \cite{Middleton:2020asl} and/or cosmological origin
including stable strings \cite{Ellis:2020ena,Blasi:2020mfx,Samanta:2020cdk,Bian:2020urb} as well as metastable strings with $\sqrt{\kappa} \simeq 8$ \cite{Buchmuller:2020lbh}. 

Given the exponential dependence of the decay rate on the parameter $\kappa$, and considering monopole masses larger than the string scale, metastable strings were generally assumed to be effectively stable \cite{Vilenkin:2000jqa}. However,
one easily verifies that a value $\sqrt{\kappa} \sim 8$ can indeed be obtained in realistic GUTs. Taking as a simple example the electroweak gauge group
$\text{SU}(2)_L\times\text{SU}(2)_R \times \text{U}(1)_{B-L}$,
broken to $\text{SU}(2)_L\times \text{U}(1)_R\times \text{U}(1)_{B-L}$ 
by an $\text{SU}(2)_R$ triplet at a scale $v_u$, and further to
$\text{SU}(2)_L\times \text{U}(1)_Y$ by an $\text{SU}(2)_R$ doublet at
scale $v_s$,
with $Y = T^3_R + \frac{1}{2}(B-L)$, one finds \cite{Buchmuller:2021dtt}
\begin{equation}\label{kappa1}
  \kappa = \frac{m_M^2}{\mu} \sim \frac{ 4 \pi}{g^2 \cos^2{\Theta}}
  \left(\frac{m_V}{m_X}\right)^2\ ,
\end{equation}
where monopole mass and string tension for isolated systems have been
used as as a rough estimate.
Here $m_V = g\sqrt{2v_u^2+v_s^2}/\sqrt{2}$ and $m_X =
gv_s/(\sqrt{2}\cos\Theta)$
are the masses of the charged and neutral vector bosons,
respectively, and $\Theta$ is the mixing angle of the $\text{U}(1)_R$ and
$\text{U}(1)_{B-L}$ vector bosons. For an embedding of the electroweak group
into a Pati-Salam or $\text{SO}(10)$ GUT group one has $\tan{\Theta} = \sqrt{3/2}$,
and with a GUT-scale gauge coupling $g\simeq 1/\sqrt{2}$ one obtains $\sqrt{\kappa} \sim 8$
for $m_V \sim m_X$, i.e., the $\text{SU}(2)$ and $\text{U}(1)$ breaking scales
have to be of the same order of magnitude.

Computing the GW spectrum emitted from a cosmic string network is a
complicated problem where many questions are still open, even for stable strings
(for a discussion and references, see \cite{Martins:2000cs,Blanco-Pillado:2017oxo,Auclair:2019wcv}).
One first studies the approach of the network to a scaling regime in
which the relative contribution to the total energy density of the
universe remains constant and computes 
the number density of non-self-interacting loops per unit string length by
means of Nambu-Goto string simulations. Together with a model for the
average power radiated off in GWs by each loop, this yields the GW spectrum.

For metastable cosmic strings a detailed study has been carried out in 
\cite{Buchmuller:2021mbb}, see also \cite{Gouttenoire:2019kij,Dunsky:2021tih}.
Compared to stable strings, the kinetic equations for the string
network are modified to take
into account the decay of string loops to segments through the
formation of a monopole--antimonopole pair, as well as the formation
of segments from longer segments and super-horizon
strings.
As demonstrated in \cite{Buchmuller:2021mbb}, the GW spectrum generated by string loops alone provides a good approximation to the full spectrum in most of the parameter space.
The key change compared to stable cosmic strings is an additional
decay term in the kinetic equation for the loop number density
accounting for the monopole--antimonopole formation on the loops.
Following \cite{Leblond:2009fq}, the number density
$\overset{\circ}{n}(\ell, t)$ of loops with length $\ell$ is matched to
the loop number density of stable strings at early times,
$t \ll t_s =  1/\Gamma_d^{1/2}$, which is determined by
numerical simulations. In the radiation-dominated era one
then obtains for the loop number
density of metastable strings at $t > 1/\Gamma_d^{1/2}$ \cite{Buchmuller:2021mbb},
\begin{align}
 \overset{\circ}{n}^\text{rad}(\ell, t) = \frac{B}{t^{3/2} (\ell + \Gamma G \mu t)^{5/2}}\, e^{- \Gamma_d [ \ell (t - t_s) + \tfrac{1}{2} \Gamma G \mu (t - t_s)^2]} \, \Theta(\alpha t_s - \ell - \Gamma \mu ( t - t_s)) \,.
 \label{eq:nmetastable}
\end{align} 
Here, the exponential factor accounts for the decay of the loops at $t
> t_s$ through the generation of monopoles, and the Heaviside function
ensures that loop formation only occurs at $t < t_s$. $\Gamma \simeq
50$ is the power in GWs emitted by a single loop, and the parameters
$B$ and $\alpha$ are determined by numerical simulations.
Expressions for the loop number densities involving evolution during
the matter-dominated era, as well as expressions for the number
densities of super-horizon strings and segments, are given in \cite{Buchmuller:2021mbb}. 

As for stable cosmic strings, a population of string loops with number
density $\overset{\circ}{n}(\ell, t)$ yields the spectral energy
density in GWs today normalized by the critical energy density
\cite{Blanco-Pillado:2017oxo,Auclair:2019wcv},
\begin{align}
 \Omega_\text{gw}(t_0, f) = \frac{16 \pi (G\mu)^2}{3 H_0^2 f} \sum_k k
  P_k \int_0^{z_i} \frac{dz'}{H(z') (1 + z')^6}\, \overset{\circ}{n}(2
  k/f', t(z')) \ .
 \label{eq:OmegaGW}
\end{align}
Here $f' = 2 k/\ell$ indicates the GW frequency, emitted by a loop of
length $\ell$ oscillating in its $k$th harmonic excitation at time
$t'$ of GW emission, and the integration over $t'$ from some initial
time $t_i$ to $t_0$ has been traded for an integration over the
redshift $z$; $H(z)$ is the Hubble parameter, and the argument of the
loop number density ensures that we are accounting for all GWs emitted
at frequency $f'$ such that after red-shifting, they are observed at
frequency $f$ today. $\Omega_{\text{gw}}$ depends on two parameters,
the string tension $G\mu$ and the decay parameter ${\kappa}$.

\begin{figure}
\centering
\includegraphics[width = 0.6 \textwidth]{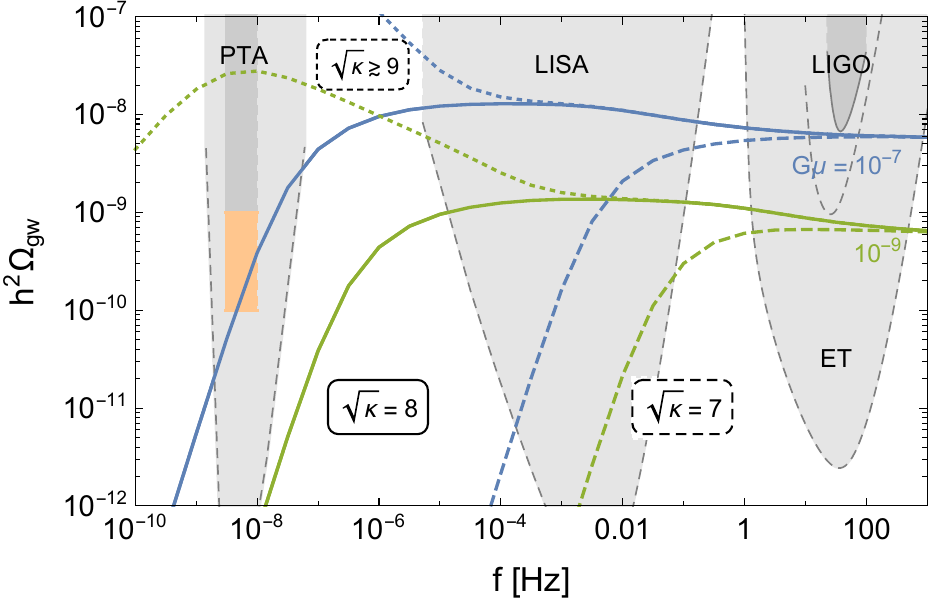}
\caption{GW spectrum from metastable cosmic strings with a string
  tension of $G\mu = 10^{-7}$ (blue) and $G\mu = 10^{-9}$
  (green), with $g_*=g_*(T)$. Different line styles indicate different string lifetimes,
  ranging from $\sqrt{\kappa} = 7$ (dashed) over $\sqrt{\kappa} = 8$
  (solid) to the limit of (quasi-)stable strings $\sqrt{\kappa}
  \gtrsim 9$ (dotted).
The gray-shaded areas indicate the sensitivity of current (solid) and
planned (dashed) GW experiments. The orange region indicates the
preferred region of the possible GWB signal observed by PTAs. See \cite{Buchmuller:2021mbb,Buchmuller:2023aus}.}
\label{fig:spectrum}
\end{figure}

Fig.~\ref{fig:spectrum} shows the GW spectrum obtained by inserting
Eq.~\eqref{eq:nmetastable} and the corresponding expression for the
matter-dominated era into Eq.~\eqref{eq:OmegaGW}. 
The colored curves indicate predictions for the GW spectrum for
different values of $\kappa$ and $\mu$. 
The dotted black curves show the limit of stable cosmic strings.
Large frequencies correspond to GWs produced at early times, and hence the spectrum produced by stable and metastable strings is identical, featuring a plateau at
\begin{align}
 \Omega_\text{gw}^\text{plateau} \simeq \frac{128 \pi}{9} B \, \Omega_r \left(\frac{G \mu}{\Gamma} \right)^{1/2} \,,
\end{align}
where $\Omega_r h^2 = 4.15 \cdot 10^{-5}$ is the density parameter of radiation today.
At small frequencies, the decay of the metastable cosmic
string loops suppresses the GW signal, leading to a drop in the
spectrum proportional to $f^2$.

\begin{figure}[t]
\centering
\includegraphics[width = 0.6 \textwidth]{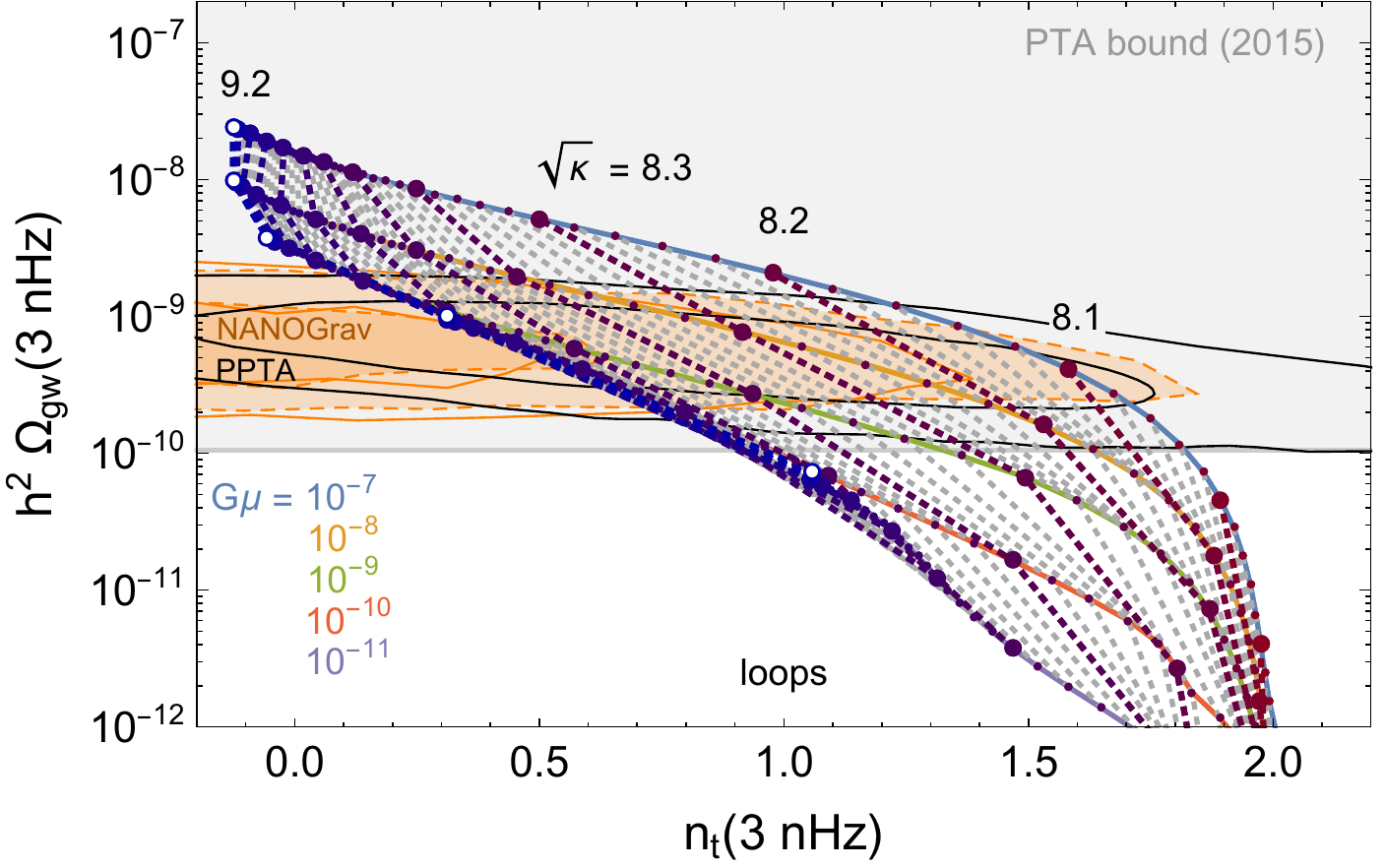}
\caption{Predictions for the GW spectrum in the frequency range of pulsar
  timing arrays, for different values of $G\mu$ and $\sqrt{\kappa}$,
  including contributions from loops decaying during the matter era.
The bound from the Parkes Pulsar Timing Array
(PPTA)~\cite{Shannon:2015ect} published in 2015  is shown in gray;
the 1/2 $\sigma$ credible regions of
NANOGrav~\cite{Arzoumanian:2020vkk}  and
PPTA~\cite{Goncharov:2021oub} are indicated in orange and
black, respectively. From \cite{Buchmuller:2021mbb}.}
\label{fig:summary_loops}
\vspace{0.5cm}
%
\includegraphics[width = 0.6 \textwidth]{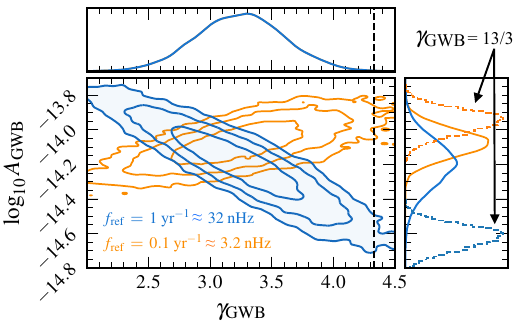}
\caption{Posterior probability distribution of GWB amplitude and
  spectral exponent in a Hellings-Downs power-law model, showing 1/2/3 $\sigma$ credible
  regions. The value $\gamma_{\text{GWB}}=13/3$ (dashed black line) is
  included in the 99\% credible region. The amplitude is referenced to
  $f_{\text{ref}} = 1\text{yr}^{-1}$ (blue) and $0.1\text{yr}^{-1}$
  (orange). The dashed blue and orange curves in the
  $\text{log}_{10}A_{\text{GWB}}$ subpanel shows its marginal posterior
  density for a $\gamma_{\text{GWB}}=13/3$ model, with 
  $f_{\text{ref}} = 1\text{yr}^{-1}$  and $0.1\text{yr}^{-1}$,
  respectively. From \cite{NANOGrav:2023gor}.}
\label{fig:NANOG2}
\end{figure}

In the PTA frequency band a power law model is used to describe the data,
\begin{equation}
\Omega_{\text{gw}} = \frac{2\pi^2f_{\text{PTA}}^2
  A^2_{\text{GWB}}}{3H_0^2}\left(\frac{f}{f_{\text{PTA}}}\right)^{n_t} \ ,
\end{equation}
where $f_{\text{PTA}}$ is a suitably chosen reference frequency. For 
specified values of $G\mu$ and ${\kappa}$, the
amplitude $h^2\Omega_{\text{gw}}$ and the spectral index $n_t$
are predicted. Lines of constant $G\mu$ and ${\kappa}$, respectively, are
displayed in Fig.~\ref{fig:summary_loops} and compared with the
NANOGrav data~\cite{Arzoumanian:2020vkk} (orange),
with the solid contours showing the
$1\sigma$ and $2\sigma$ regions.
The black solid lines show the preferred region reported by
PPTA~\cite{Goncharov:2021oub}  when performing a similar analysis.
The $2\sigma$ regions are consistent with the parameter range
$10^{-11} \lesssim G \mu \lesssim 10^{-7}$ and $\sqrt{\kappa} \gtrsim
8$. Note that the PTA data are
essentially a measurement of the amplitude with the spectral index still subject
to a large uncertainty and largely uncorrelated with the amplitude.

Pulsar timing array observations have entered a new phase
with evidence for  Hellings--Downs angular correlation, the
smoking-gun signal of a GWB, which has recently been reported by PTA collaborations
across the
world~\cite{NANOGrav:2023gor,Antoniadis:2023ott,Reardon:2023gzh,Xu:2023wog}. 
The result of the NANOGraph collaboration for amplitude and spectral index is
shown for two reference frequencies in Fig.~\ref{fig:NANOG2}, where
instead of $h^2 \Omega_{\text{gw}}$ and $n_t$ the variables
$A_{\text{GWB}}$ and $\gamma_{\text{GWB}} = 5 -n_t$ are used.
The observed amplitude is consistent with the result reported in 
\cite{Arzoumanian:2020vkk}, $10^{-10} \lesssim h^2\Omega_\text{gw}
\lesssim 10^{-9}$ at the PTA peak sensitivity $f_\text{PTA} = 3$~nHz.
The preferred spectral index is now restricted to $0 \lesssim n_t
\lesssim 3$, with the PPTA data set \cite{Reardon:2023gzh} preferring slightly
smaller values and  the EPTA 10.5 year data set
\cite{Antoniadis:2023ott} preferring larger values. 

Beyond the astrophysical interpretation of the GWB in terms of
inspiraling supermassive black-hole
binaries~\cite{Middleton:2020asl,NANOGrav:2023hfp,Antoniadis:2023xlr}
there are several viable cosmological interpretations including cosmic strings
(for an overview, see, e.g.~
\cite{Madge:2023cak,NANOGrav:2023hvm,Antoniadis:2023xlr}).
As consequence of the restricted range of the spectral index,
metastable strings are now favoured over stable strings
\cite{NANOGrav:2023hvm}. Cosmic superstrings
\cite{Dvali:2003zj,Copeland:2003bj,Ellis:2023tsl},
with a GW spectrum estimated by a rescaled GW spectrum of stable strings,
provide an even better fit to the data \cite{NANOGrav:2023hvm}.
Upcoming data will improve our understanding of the spectral index,
the isotropy and the presence of resolvable individual sources in this
GWB, which will all help to distinguish an astrophysical from a
cosmological origin, as well as different cosmic string interpretations.

For the physical interpretation of the PTA results it is useful to
express the decay parameter $\kappa$ in terms of the $\text{U}(1)$ and $\text{SU}(2)$
symmetry breaking scales $v_s$ and $v_u$, respectively. The string
tension is given by $\mu \simeq 2\pi v^2_s$, and a monopole with flux
$2\pi n/g$ has mass $m_M \simeq 2\pi n v_u/g$.
For a monopole--string--antimonopole
configuration, the magnetic fluxes of the string and the
(anti)monopole have to match (for a discussion, see,
e.g.~\cite{Kibble:2015twa}).
The string solution with lowest energy has winding number $n=1$.
For the considered Pati-Salam embedding the symmetry breaking doublet
has $\text{U}(1)$ charge $1/2$ and carries magnetic
flux $4\pi/g$. This can be matched by a $n=2$ monopole with mass $m_M \simeq
4\pi v_u/g$, which yields for the decay parameter\footnote{From
  Eq.~\eqref{kappa1} one obtains
$\kappa = (4\pi/g^2)(2v_u^2 + v_s^2)/v_s^2$. For $v_u \gg v_s$, this
agrees with Eq.~\eqref{kappa2}, a consequence of the fact that in Eq.~\eqref{kappa1}
the mass of a 't Hooft-Polyakov monopole was used
which has charge $n=2$. The difference between the expressions 
\eqref{kappa1} and \eqref{kappa2} in the regime $v_u \sim v_s$ illustrates the
theoretical uncertainties of the two estimates.}\cite{Buchmuller:2023aus}
\begin{equation}\label{kappa2}
  \kappa = \frac{m_M^2}{\mu} \sim \frac{8\pi}{g^2}\frac{v_u^2}{v_s^2} \,.
\end{equation}
With $g^2 \sim 1/2$ at the unification scale, one finds again
 that $\sqrt{\kappa} \sim 8$ can be achieved if the two symmetry breaking scales $v_u$
 and $v_s$ have comparable size. Note that this value of $\kappa$
 favours supersymmetric GUTs. Smaller gauge couplings in
 non-supersymmetric GUTs, like $g^2 \sim 1/4$ at the GUT scale, would
 require $v_u < v_s$ and may render the cosmic strings unstable.
 
The decay rate \eqref{decayrate} is derived for infinitely thin strings
with point-like monopoles at the end. Clearly, this approximation is
very questionable in the case $v_u \sim v_s$. Moreover, the effect of
scalar fields has been neglected in the expressions \eqref{kappa1} and
\eqref{kappa2} for the decay parameter, which for comparable symmetry
breaking scales can be rough estimates at best. It is not even clear
whether in this parameter range metastable strings exist at all. We
know that for $v_u \rightarrow 0$ unstable dumbbells form instead
of metastable strings \cite{Buchmuller:2021dtt}. Hence, at some critical
value of the ratio $v_u/v_s$ the domain of metastabily should turn
into a domain of instability. The difficult problem of computing the
decay rate of metastable strings beyond the thin-defect approximation
has recently been addressed in \cite{Chitose:2023dam}, building on
earlier work in \cite{Shifman:2002yi}. Starting vom the ``unwinding
process'' discussed in \cite{Shifman:2002yi}, a bounce action
describing the decay of metastable strings has been obtained and
numerically analyzed in the regime $v_u \sim v_s$. The results show
that for scalar masses of the same order of magnitude there is indeed
a regime of metastability with $\sqrt{\kappa} \sim 8$, as required by
the PTA data \cite{Chitose:2023dam}.

\begin{figure}[t]
\centering
\includegraphics[width = 0.7 \textwidth]{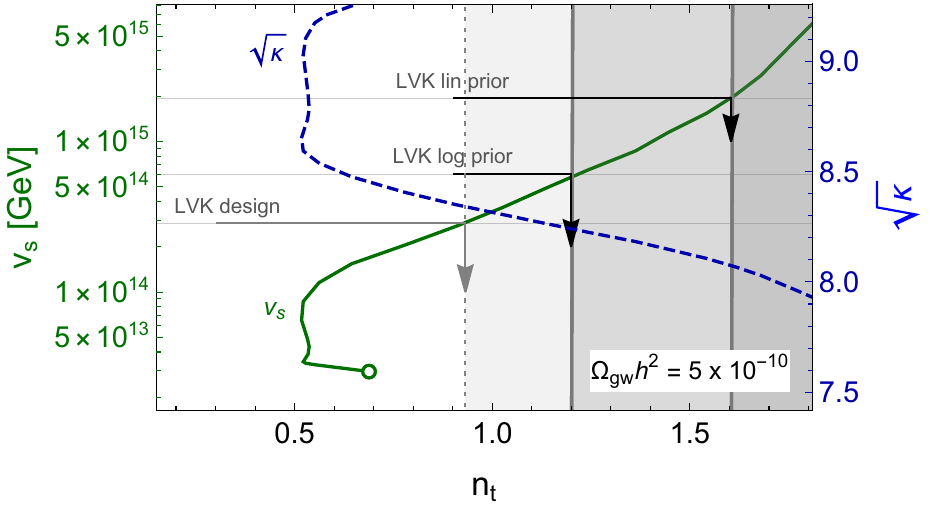} \hfill
\caption{$U(1)$ symmetry breaking scale $v_s$ (solid green, left axis)
  and decay parameter $\sqrt{\kappa}$ 
  (dashed blue, right axis) as functions of the spectral
  index $n_t$ of the GWB, assuming an amplitude
  $h^2\Omega_\text{gw} \simeq 5 \cdot 10^{-10}$ 
  at $f_{\text{PTA}} = 3$~nHz. The grey region indicates values
  $G\mu > 1.5 \cdot 10^{-8}$ which is disfavoured by
  LIGO-Virgo--KAGRA (LVK) for a logarithmic (linear) prior on the
  amplitude of the GWB signal in the LVK
  band~\cite{KAGRA:2021kbb}; the light grey region corresponds to the
  design sensitivity.  From \cite{Buchmuller:2023aus}.
}
\label{fig:nt}
\end{figure}

Using Eq.~\eqref{kappa2} and $\mu \simeq 2\pi v^2_s$, the two
observables $\Omega_{\text{gw}}$ and $n_t$ allow to determine the two
symmetry breaking scales, $v_u$ and $v_s$. As shown in
Fig.~\ref{fig:summary_loops}, the amplitude $h^2\Omega_\text{gw}$ 
is rather well determined at the reference frequency $f_{\text{PTA}} = 3$~nHz,
whereas the spectral index $n_t$ is much more uncertain. This
allows a simplified analysis:
Fixing the amplitude of the GW spectrum at $3$~nHz to the
representative value
$h^2\Omega_\text{gw} = 5 \cdot 10^{-10}$, the analysis in
\cite{Buchmuller:2021mbb} can be used to determine $G\mu$ and
$\sqrt{\kappa}$ as functions of $n_t$ (see
Fig.~\ref{fig:summary_loops}), which can be translated to functions
$v_u$ and $v_s$ of $n_t$. The results for $\sqrt{\kappa}$ and $v_s$
are shown in Fig.~\ref{fig:nt}.
As the string lifetime and hence $\kappa$ is increased, the
string tension $G\mu$ needs to be reduced to maintain the same GWB
amplitude at 3~nHz. An increase in $\kappa$
comes with a decrease in $n_t$ at $f_{\text{PTA}}$, until with a further increase of
$\kappa$ the $f^2$ part of the spectrum enters the PTA band and the
spectral index starts increasing again.
In the limit of stable strings, $\sqrt{\kappa} \rightarrow \infty$,
one finds $G\mu \simeq 4 \cdot 10^{-11}$ and $n_t \simeq 0.7$.
For small $\kappa$
(large $v_s$),
the GWB will exceed the bound
$\Omega_\text{gw} \leq 5.8 \cdot 10^{-9}$ set by the
LIGO--Virgo--KAGRA (LVK) collaboration in the
100~Hz range for a logarithmic prior~\cite{KAGRA:2021kbb}. As shown in
Fig.~\ref{fig:nt}, this
leads to an upper bound on the spectral index, $n_t \lesssim 1.2$.
From Fig.~\ref{fig:NANOG2} one reads off the 2$\sigma$ upper bound
$\gamma_{\text{GWB}} \lesssim 4.1$, corresponding to the 2$\sigma$
lower bound on the spectral index $n_t \gtrsim 0.9$, which
coincides with the LVK design sensitivity for setting an upper bound
on $n_t$.

\section{GUT model building}

Metastable strings are a characteristic prediction of GUTs which lead,
via several steps of spontaneous symmetry breaking, to the Standard
Model gauge group
$\text{G}_{SM} =
\text{SU}(3)_C\times\text{SU}(2)_{L}\times\text{U}(1)_{Y}$.
Strings with tensions above the electroweak scale result from the
breaking of a $\text{U}(1)$ group that commutes with
$\text{G}_{SM}$. Similarly, monopoles arise from the breaking of a
non-Abelian gauge group that leaves a $\text{U}(1)$ subgroup
unbroken. If this $\text{U}(1)$ overlaps with the $\text{U}(1)$ symmetry
responsible for string formation, the strings become metastable
by quantum tunneling. Hence, gauge groups containing the SM 
which can give rise to metastable strings must have at least rank $5$.
A typical sequence of SM embeddings is
\begin{equation}\label{embLR}
  \begin{split}
    \text{G}_{SM}
  \subset \text{SU}(3)_C\times\text{SU}(2)_{L}\times\text{SU}(2)_{R}
  \times\text{U}(1)_{B-L} 
   \subset
  \text{G}_{PS}\subset \text{SO}(10) \subset \text{E}(6) \subset \dots \,,
\end{split}
\end{equation}
where $\text{G}_{PS} = \text{SU}(4)\times\text{SU}(2)_{L}
\times\text{SU}(2)_{R}$ denotes the Pati--Salam group.
  
If a symmetry group $\text{G}$  is broken to a subgroup
$\text{H}$, the quotient $\mathcal{M} = \text{G}/\text{H}$
corresponds to the manifold of degenerate vacuum states.
The types of defects that may be formed in the symmetry breaking are
governed by the topology of $\mathcal{M}$, which is encoded in the
homotopy groups $\pi_n(\mathcal{M})$ (for a review and referenes, see,
e.g. \cite{Vilenkin:2000jqa}. Topologically stable
strings can form if the first homotopy group is nontrivial,
$\pi_1(\mathcal{M})  \neq I$, and topologically stable magnetic
monopoles can arise if the
second homotopy group is nontrivial, $\pi_2(\mathcal{M}) \neq I$.
In two-step symmetry breakings
$\text{G} \rightarrow \text{H} \rightarrow \text{K}$, where the
homotopy group $\text{G}/\text{K}$ is trivial, but the homotopy groups
of the individual steps, $\text{G}/\text{H}$ and
$\text{H}/\text{K}$, are nontrivial, metastable defects can form.

The vacuum manifold crucially depends on the
chosen Higgs representation.
A simple example is the breaking of $\text{SO}(10)$ to the Standard
Model group via $\text{SU}(5)$ \cite{Kibble:1982ae}. The breaking chain
\begin{equation}\label{SU5A}
  \text{SO}(10) \stackrel{\bf{45}}{\rightarrow} \text{SU}(5) \times
  \text{U}(1) \stackrel{\bf{45}\oplus\bf{126}}{\rightarrow}
  \text{G}_{SM} \times \mathbb{Z}_2
\end{equation}
yields stable monopoles and, in the second step, also stable strings.
On the contrary, for the closely related symmetry
breaking with a $\bf{16}$-plet,
\begin{equation}\label{SU5B}
  \text{SO}(10) \stackrel{\bf{45}}{\rightarrow} \text{SU}(5) \times
  \text{U}(1) \stackrel{\bf{45}\oplus\bf{16}}{\rightarrow}
  \text{G}_{SM} \,,
\end{equation}
the homotopy group of $\mathcal{M} = \text{SO}(10)/\text{G}_{SM}$ is trivial, $\pi_1(\mathcal{M})  = I$, and there
are no topologically stable strings. However, 
cosmologically interesting metastable strings can now form.

Realistic GUTs require large Higgs representations in order to break
the GUT gauge group down to the SM, which complicates the
analysis of symmetry breaking patterns. Moreover, nonsupersymmetric
models are sensitive to large
radiative corrections and hence suffer from a severe naturalness
problem. This lead to even more complicated models: supersymmetric GUTs with even
more complicated Higgs sectors. As a possible alternative, one can
consider higher-dimensional theories such as orbifold GUTs or string
models (for a review, see,
e.g.~\cite{Raby:2017ucc,Hebecker:2021egx}).
In these constructions,
the GUT gauge group is first partially broken in a
geometric way, by the compactification of extra
dimensions, and only the remnant subgroup remaining after this first
step is further reduced to the SM group via conventional spontaneous
symmetry breaking.

The simplest group leading to potentially realistic metastable strings is
$\text{SU}(3)_C\times\text{SU}(2)_{L}\times\text{SU}(2)_{R}$, the
electroweak subgroup of the Pati--Salam and $\text{SO}(10)$ GUT
groups. A Higgs triplet $U$, breaking $\text{SU}(2)_{R}$ to
$\text{U}(1)_{R}$ with $\langle U^a\rangle = v_u\delta_{a3}/\sqrt{2}$,
is contained in the adjoint representation, and Higgs doublets 
$S$ and $S_c$, breaking $\text{SU}(2)_{R}\times U(1)_{R}$ to $\text{U}(1)_{Y}$
with $\langle S_i\rangle =\langle S_{ci}\rangle = v_s\delta_{i1}/\sqrt{2}$,
are contained in the Pati--Salam multiplets
$\left(\mathbf{4},1,\mathbf{2}\right)$ and
$\left(\mathbf{\bar{4}},1,\mathbf{\bar{2}}\right)$, and
the  $\text{SO}(10)$ multiplets $\mathbf{16}$ and
$\mathbf{\overline{16}}$,  respectively.
Embedding the doublets $S$, $S_c$ in $\mathbf{16}$-, $\mathbf{\overline{16}}$-plets
  $\Phi$, $\Phi^c$ of $\text{SO}(10)$
  implies that heavy Majorana neutrino masses must
be generated by the nonrenormalizable operator
\begin{equation}
  \mathcal{L}_n = \frac{1}{M_*}\,h_{ij}\, S^T L^c_i S^T L^c_j \subset
\frac{1}{M_*}\, h_{ij}\, \Phi^c \psi_i {\Phi^c} \psi_j \,.
\end{equation}
  Here, the fields $L^c_i = (n^c_i,e^c_i)^T$,
  $i=1,..,3$, denote the $\text{SU}(2)_R$ doublets of right-handed neutral and charged
  leptons that are contained in the $\text{SO}(10)$ $\mathbf{16}$ representations
  $\psi_i$ of matter,
  and $h_{ij}$ are Yukawa couplings. Alternatively, one can 
  break $\text{SO}(10)$ with $\mathbf{126}$-,
  $\mathbf{\overline{126}}$-plets $\tilde{\Phi}$, $\tilde{\Phi}^c$ containing
  the $\text{SU}(5)$ singlets $\tilde{S}$, $\tilde{S_c}$. Heavy
  neutrino masses are now generated by the renormalizable
  couplings\footnote{This was assumed in
    \cite{Buchmuller:2012wn}. Note that string compactifications
    prefer symmetry breaking with $\mathbf{16}$- and
    $\mathbf{\overline{16}}$-plets which, contrary to $\mathbf{126}$-
and $\mathbf{\overline{126}}$-plets, are contained in the adjoint
representation of $\text{E}_8$ \cite{Raby:2017ucc,Hebecker:2021egx}.}   
  \begin{equation}
  \mathcal{L}_n = h_{ij} \tilde{S} L^c_i L^c_j \subset
  h_{ij} \tilde{\Phi} \psi_i \psi_j \ .
\end{equation}
The VEVs of $\tilde{S}$, $\tilde{S_c}$ leave a $\mathbb{Z}_2$ discrete 
symmetry unbroken, which leads to topologically stable strings.
Various aspects of metastable strings in $\text{SO}(10)$ models have
been discussed in
\cite{Buchmuller:2019gfy,Buchmuller:2021dtt,Buchmuller:2023aus,Antusch:2023zjk,Fu:2023mdu,Afzal:2023cyp,Ahmed:2023pjl,Ahmed:2023rky}.
A viable alternative are flipped-$\text{SU(5)}$ models
\cite{Lazarides:2023rqf,King:2023wkm}. 

Metastable strings that can account for the observed PTA GWB require a two-step
GUT symmetry breaking whose scales $v_u$ and $v_s$ satisfy the conditions
\begin{equation}
  v_u \sim v_s \sim \text{few} \times 10^{14}~\text{GeV} \ll
  v_{\text{GUT}} \sim 10^{16}~\text{GeV}\ .
\end{equation}
The inequality could be explained if the symmetry breaking scales were
related to the compactification scale of a 5-dimensional
$\text{SO}(10)$ GUT rather than parameters of a 4-dimensional supersymmetric GUT. In such
higher-dimensional theories also topological defects can occur, which
may shed some light on the connection between the two symmetry
breaking scales (for a discussion and references, see,
e.g.~\cite{Dermisek:2002ri,Kim:2002im,Raby:2017ucc}).
On the observational side, it is intriguing that the PTA lower bound on
the spectral index $n_t$ coincides with the upper bound on $n_t$ corresponding
to the LVK design sensitivity (see Fig.~\ref{fig:nt}). This suggests
that a GWB may soon be
observed by the LIGO-Virgo--KAGRA collaboration.\\

\vspace{0.5cm}

\noindent
{\bf\large Acknowledgments}\\

This article is based on work with V.~Domcke, H.~Murayama and K.~Schmitz
whom I thank for an eventful collaboration. Special thanks also to
E.~Dudas for discussions on metastable strings in grand unified models.
I am grateful to the organizing committee for arranging  this international conference
dedicated to Valery Rubakov in Yerevan in difficult times.

\newpage

\bibliographystyle{JHEP}
\bibliography{nanoGUT}


\end{document}